\documentclass[letterpaper,titlepage,11pt]{article}
\usepackage{hyperref}

\usepackage{amssymb,amsmath,amsfonts}
\usepackage{epsfig}

\def\clock{{\count0=\time
           \divide\count0 60
           \ifnum\count0<10 0\fi\the\count0
           \multiply\count0 -60 \advance\count0 \time
           :\ifnum\count0<10 0\fi \the\count0
         }}
\newcommand{\timestamp}{{\small\vbox{\hbox{\tt\jobname.tex}
\hbox{\the\day/\the\month/\the\year, \clock}}}}

\newcommand{\CH}{\mathcal{H}}

\newcommand{\CN}{\mathcal{N}}

\newcommand{\Z}{\mathbb{Z}}

\newcommand{\R}{\mathbb{R}}

\newcommand{\spa}{\ , \ \ }

\newcommand{\ds}{\displaystyle}

\newcommand{\tr}{\mathop{{\rm Tr}}}

\newcommand{\ads}{\mbox{AdS}}

\newcommand{\zeth}{\zeta \Big( \frac{3}{2} \Big)}

\numberwithin{equation}{section}
\begin{document}

\begin{titlepage}

\rightline{\vbox{\small\hbox{\tt hep-th/0701088} }}
 \vskip 2.7cm

\centerline{\LARGE \bf The Hagedorn temperature in a decoupled
sector of AdS/CFT} \vskip 2cm

\centerline{\large {\bf Troels Harmark$\,^{1}$}, {\bf Kristjan R.\
Kristjansson$\,^{2}$} and {\bf Marta Orselli$\,^{2}$} }

\vskip 0.6cm

\begin{center}
\sl $^1$ The Niels Bohr Institute  \\
\sl  Blegdamsvej 17, 2100 Copenhagen \O , Denmark \\
\vskip 0.4cm
\sl $^2$ Nordita \\
\sl  Blegdamsvej 17, 2100 Copenhagen \O , Denmark \\
\end{center}
\vskip 0.6cm

\centerline{\small\tt harmark@nbi.dk, kristk@nordita.dk,
orselli@nbi.dk}

\vskip 1.5cm

\centerline{\bf Abstract} \vskip 0.2cm \noindent We match the
Hagedorn/deconfinement temperature of planar $\CN=4$ super
Yang-Mills (SYM) on $\R \times S^3$ to the Hagedorn temperature of
string theory on $\ads_5 \times S^5$. The match is done in a
near-critical region where both gauge theory and string theory are
weakly coupled. On the gauge theory side we are taking a decoupling
limit found in Ref.~\cite{Harmark:2006di} in which the physics of
planar $\CN=4$ SYM is given exactly by the ferromagnetic $XXX_{1/2}$
Heisenberg spin chain. We find moreover a general relation between
the Hagedorn/deconfinement temperature and the thermodynamics of the
Heisenberg spin chain. On the string theory side, we identify the
dual limit which is taken of string theory on a maximally symmetric
pp-wave background with a flat direction, obtained from a Penrose
limit of $\ads_5 \times S^5$. We compute the Hagedorn temperature of
the string theory and find agreement with the Hagedorn/deconfinement
temperature computed on the gauge theory side. Finally, we discuss a
modified decoupling limit in which planar $\CN=4$ SYM reduces to the
$XXX_{1/2}$ Heisenberg spin chain with an external magnetic field.

%\vskip 0.5cm \leftline{\timestamp}

\end{titlepage}

\pagestyle{plain} \setcounter{page}{1}

%\tableofcontents

%%%%%%%%%%%%%%%%%%%%%%%%%%%%%%%%%%%%%%%%%%%%%%%%%%%%%%%%%%%%%%
\section{Introduction}
\label{sec:intro}

The most beautiful example of the relation between gauge theories
and string theories is the AdS/CFT correspondence which asserts an
exact duality between $SU(N)$ $\CN = 4$ super Yang-Mills (SYM) on
$\R \times S^3$ and type IIB string theory on $\ads_5 \times S^5$
\cite{Maldacena:1997re,Gubser:1998bc,Witten:1998qj}. The AdS/CFT
correspondence is a strong/weak coupling duality. This is the power
of the correspondence but it also makes it difficult to verify its
validity. Many of the checks have involved computing physical
quantities on the gauge theory side, such as the expectation value
of Wilson loops \cite{Erickson:2000af,Drukker:2000rr} or the
anomalous dimensions of gauge theory operators
\cite{Berenstein:2002jq}, and extrapolating the results to strong
coupling in order to compare with string theory.

In this talk we instead check the validity of the AdS/CFT
correspondence avoiding the extrapolation of the results to strong
coupling, following the papers
\cite{Harmark:2006di,Harmark:2006ta,Harmark:2006ie}. The strategy we
use is to compute the Hagedorn/deconfinement temperature for planar
$\CN = 4$ SYM on $\R \times S^3$ at weak coupling $\lambda \ll 1$ in
a certain near-critical region found in \cite{Harmark:2006di} and to
match this to the Hagedorn temperature computed in weakly coupled
string theory on $\ads_5 \times S^5$, in the corresponding dual
near-critical region. This successful match mostly relies on the
matching of the low energy spectra of the gauge theory and the
string theory in the near-critical region.

In
\cite{Witten:1998zw,Sundborg:1999ue,Polyakov:2001af,Aharony:2003sx}
a relation is conjectured between the Hagedorn/deconfinement
temperature of planar $\CN = 4$ SYM on $\R \times S^3$ and the
Hagedorn temperature of string theory on $\ads_5 \times S^5$. This
is due to the discovery of a confinement/deconfinement phase
transition in planar $\CN = 4$ SYM on $\R \times S^3$ at weak
coupling $\lambda \ll 1$ \cite{Witten:1998zw}. In particular, at
high energies the theory has a Hagedorn density of states, with the
Hagedorn temperature being equal to the deconfinement temperature
\cite{Sundborg:1999ue,Polyakov:2001af,Aharony:2003sx}.

However, the fact that we do not know how to quantize string theory
on $\ads_5\times S^5$ means that we cannot directly test this
conjecture. One hope comes from considering certain Penrose limits
where the $\ads_5 \times S^5$ background becomes a maximally
supersymmetric pp-wave background
\cite{Berenstein:2002jq,Bertolini:2002nr} where type IIB string
theory can be quantized. In this case in fact the Hagedorn
temperature has been computed \cite{Greene:2002cd,
PandoZayas:2002hh}. However, in order to obtain the correspondence
with string theory, it is necessary to consider a strong coupling
limit on the gauge theory side so that most of the gauge theory
operators decouple keeping only those dual to the string states.

In this paper we take a different route by taking a decoupling limit
corresponding to being in a certain near-critical region. Using this
we find a gauge-theory/pp-wave correspondence appropriate for
verifying the relation between the Hagedorn/deconfinement
temperature of planar $\CN = 4$ SYM on $\R \times S^3$ and the
Hagedorn temperature of string theory on $\ads_5 \times S^5$.

We expect more generally that our decoupling limits can be used to
study the thermodynamics of $\CN = 4$ SYM on $\R \times S^3$ and its
string theory dual also above the Hagedorn temperature. It could in
particular be interesting to study the connection to black holes in
$\ads_5\times S^5$
\cite{AlvarezGaume:2006jg}.%
\footnote{See also \cite{Harmark:1999xt} for a related study of
black holes with R-charged chemical potentials.}

%%%%%%%%%%%%%%%%%%%%%%%%%%%%%%%%%%%%%%%%%%%%%%%%%%%%%%%%%%%%%%%%%%%%%%

\section{Gauge theory side}

The solution to find the appropriate gauge-theory/pp-wave
correspondence comes from a recently found decoupling limit of
thermal $SU(N)$ $\CN=4$ SYM on $\R \times S^3$ \cite{Harmark:2006di}
which is given by
\begin{equation}
\label{ourlim} T\rightarrow 0\spa \Omega\rightarrow 1\spa
\lambda\rightarrow 0\spa \tilde{T}\equiv\frac{T}{1-\Omega}\
\mbox{fixed}\spa \tilde{\lambda}\equiv\frac{\lambda}{1-\Omega}\
\mbox{fixed} \spa N \ \mbox{fixed}
\end{equation}
where $T$ is the temperature for $\CN=4$ SYM, $\Omega$ is the
chemical potential associated to the three R-charges $J_1, J_2,J_3$
for the $SU(4)$ R-symmetry and it is defined as
$(\Omega_1,\Omega_2,\Omega_3)=(\Omega,\Omega,0)$. $\lambda=g_{\rm
YM}^2N/4\pi^2$ is the 't Hooft coupling. In the limit \eqref{ourlim}
only the states in the $SU(2)$ sector survive, and $SU(N)$ $\CN=4$
SYM on $\R \times S^3$ reduces to a quantum mechanical theory with
temperature $\tilde{T}$ and coupling $\tilde{\lambda}$. In fact,
consider the thermal partition function of $SU(N)$ $\CN=4$ SYM on
$\R \times S^3$ with non-zero chemical potentials
\begin{equation}
\label{genZ} Z(\beta,\Omega_i) = \tr \left( e^{-\beta D + \beta
\sum_{i=1}^3 \Omega_i J_i } \right)
\end{equation}
where $\beta=1/T$ is the inverse temperature, $D$ is the dilatation
operator and the trace is taken over all gauge invariant states,
corresponding to all the multi-trace operators. It is convenient to
combine the R-charges $J_1$ and $J_2$ into the following charges
\begin{align}
\label{jsz} J \equiv J_1 + J_2, \qquad S_z \equiv \frac{1}{2}
(J_1-J_2).
\end{align}
At weak coupling and in the decoupling limit \eqref{ourlim} the
partition function \eqref{genZ} reduces to
\begin{equation}
\label{Zsu2} Z(\tilde{\beta}) = {\tr}_{\CH} \left( e^{-\tilde{\beta}
H} \right)
\end{equation}
with $H$ being the Hamiltonian $ H = D_0 + \tilde{\lambda} D_2$ and
$\tilde{\beta}=1/\tilde{T}$. We see that $SU(N)$ $\CN=4$ SYM on $\R
\times S^3$ in the limit \eqref{ourlim} reduces to a quantum
mechanical theory with Hilbert space $\CH$ given by the $SU(2)$
sector. $\tilde{T}$ and $\tilde{\lambda}$ can be regarded as the
effective temperature and coupling of the theory.

Moreover, in the planar limit $N=\infty$,  $\tilde{\lambda}D_2$
becomes the Hamiltonian for the ferromagnetic $XXX_{1/2}$ Heisenberg
spin chain (without magnetic field) where~\cite{Minahan:2002ve}
\begin{equation}
D_2 = \frac{1}{2} \sum_{i=1}^L ( I_{i,i+1} - P_{i,i+1} )
\end{equation}
for a chain of length $L$, where $I_{i,i+1}$ and $P_{i,i+1}$ are the
identity operator and the permutation operator acting on letters at
position $i$ and $i+1$. We can therefore write the single-trace
partition function as \cite{Harmark:2006di}
\begin{equation}
\label{ZST} Z_{\rm ST} (\tilde{\beta}) = \sum_{L=1}^\infty
e^{-\tilde{\beta} L} Z^{\rm (XXX)}_L ( \tilde{\beta} )
\end{equation}
where
\begin{equation}
\label{ZXXX} Z^{\rm (XXX)}_L ( \tilde{\beta} ) = {\tr}_L \left(
e^{-\tilde{\beta}\tilde{\lambda} D_2} \right)
\end{equation}
is the partition function for the ferromagnetic $XXX_{1/2}$
Heisenberg spin chain of length $L$. Note that ${\tr}_L$ here refers
to the trace over single-trace operators with $J=L$ in the $SU(2)$
sector. The spin chain is required to be periodic and
translationally invariant in accordance with the cyclic symmetry of
single-trace operators. Using the standard relation between the
single-trace and multi-trace partition functions, we get
\cite{Harmark:2006di}
\begin{equation}
\label{planarZ} \log Z (\tilde{\beta}) = \sum_{n=1}^\infty
\sum_{L=1}^\infty \frac{1}{n} e^{- \tilde{\beta} nL} Z^{\rm (XXX)}_L
( n \tilde{\beta} )
\end{equation}
Therefore, the partition function of planar $SU(N)$ $\CN=4$ SYM on
$\R \times S^3$ in the decoupling limit \eqref{ourlim} is given
exactly by \eqref{planarZ} from the partition function $Z^{\rm
(XXX)}_L(\tilde{\beta})$ of the ferromagnetic $XXX_{1/2}$ Heisenberg
spin chain \cite{Harmark:2006di}. Using this interesting result we
obtain a direct connection between the Hagedorn/deconfinement
temperature for finite $\tilde{\lambda}$ and the thermodynamics of
the Heisenberg spin chain expressed by the
relation~\cite{Harmark:2006ta}
\begin{equation}
\label{rel} \tilde{T}_H = \frac{1}{V\!
\big(\tilde{\lambda}^{-1}\tilde{T}_H\big)}
\end{equation}
where $t=\tilde{\lambda}^{-1}\tilde{T}_H$ is the temperature for the
ferromagnetic Heisenberg chain with Hamiltonian $D_2$ and $-tV(t)$
is the thermodynamic limit of the free energy per site for the
Heisenberg chain. The previous relation can be used to compute the
Hagedorn temperature as a function of the coupling $\tilde \lambda$.
In the large $\tilde{\lambda}$ limit\footnote{The small $\tilde
\lambda$ regime is related to the high temperature limit of the
Heisenberg model and it is analyzed in \cite{Harmark:2006ta}.} the
Hagedorn temperature corresponds to the low temperature limit of the
Heisenberg chain, and we obtain~\cite{Harmark:2006ta}
\begin{equation}
\label{ourhag} \tilde{T}_H = (2\pi)^{\frac{1}{3}} \left[ \zeta \Big(
\frac{3}{2} \Big) \right]^{-\frac{2}{3} }
\tilde{\lambda}^{\frac{1}{3}}
\end{equation}
where $\zeta(x)$ is the Riemann zeta function. Note that the low
energy behavior of the Heisenberg chain, and thereby of the gauge
theory, is tied to the large $\tilde{\lambda}$ limit. In this region
the dominant states for the $D_2$ Hamiltonian are the low energy
states of the Heisenberg spin chain. In fact, the low energy
spectrum consisting of the chiral primary vacua with the magnon
spectrum gives rise to the Hagedorn temperature \eqref{ourhag}. In
the next section we will show that the same result \eqref{ourhag}
can be obtained by a direct string theory computation.

Before moving to the string theory side, we want to comment on a
more general situation where we study thermal $SU(N)$ $\CN=4$ SYM on
$\R \times S^3$ with chemical potentials for the R-charges for the
$SU(4)$ R-symmetry taken to be
$(\Omega_1,\Omega_2,\Omega_3)=(\Omega+h,\Omega-h,0)$. We see that
for $h=0$ we have $\Omega_1=\Omega_2=\Omega$ as previously
considered. In this new situation the decoupling limit is given by
\cite{Harmark:2006ie}
\begin{equation}
\label{hlim} \Omega\rightarrow 1,\quad
\tilde{T}\equiv\frac{T}{1-\Omega}\ \mbox{fixed},\quad
\tilde{h}\equiv\frac{h}{1-\Omega}\ \mbox{fixed},\quad
\tilde{\lambda}\equiv\frac{\lambda}{1-\Omega}\ \mbox{fixed}, \quad N
\ \mbox{fixed}
\end{equation}
The partition function can be written as
\begin{align}
\label{decpartfct} Z(\tilde \beta,\tilde h) = {\tr}_\mathcal{H}
\left(e^{-\tilde\beta H}\right)
\end{align}
where the decoupled Hamiltonian $H = D_0 + \tilde \lambda D_2 -
2\tilde h S_z$ is the Hamiltonian for the ferromagnetic $XXX_{1/2}$
Heisenberg model in the presence of an external magnetic field of
magnitude $\tilde h$. The trace is again restricted to the $SU(2)$
sector. Our new decoupling limit \eqref{hlim} generalizes the limit
\eqref{ourlim} found in \cite{Harmark:2006di}. In fact it reduces to
that for $\tilde{h}=0$. We can in principle compute the full
partition function \eqref{decpartfct} for any value of
$\tilde{\lambda}$ and $\tilde{h}$. We thus have an extra parameter
$\tilde{h}$ that can be regarded both as a magnetic field, and also
as an effective chemical potential.

To compute the Hagedorn temperature we then use the relation
\eqref{rel} and we obtain~\cite{Harmark:2006ie}
\begin{equation}
\label{hhag} \tilde{T}_H = \frac{ (2\pi)^{\frac{1}{3}} ( 1-\tilde
h)^{\frac{2}{3}}} { \zeta(\frac{3}{2})^{\frac{2}{3}} }
\tilde{\lambda}^{\frac{1}{3}}+\frac{4(2\pi)^{\frac{2}{3}}\sqrt{\tilde
h}(1-\tilde h)^{\frac{1}{3}}}{3\, \zeta(\frac{3}{2})^{\frac{4}{3}}}
\,\tilde{\lambda}^{\frac{1}{6}}+\mathcal{O}(\tilde\lambda^0).
\end{equation}
%

%%%%%%%%%%%%%%%%%%%%%%%%%%%%%%%%%%%%%%%%%%%%%%%%%%%%%%%%%%%%%%%%%%%%

\section{String theory side}

Using the AdS/CFT correspondence, we find the following decoupling
limit of string theory on $\ads_5\times S^5$, dual to the limit
\eqref{ourlim},
\begin{equation}
\label{ouradslim} \epsilon \rightarrow 0 \spa \tilde{H} \equiv
\frac{E-J}{\epsilon} \ \mbox{fixed} \spa \tilde{T}_{\rm str} \equiv
\frac{T_{\rm str}}{\sqrt{\epsilon}} \ \mbox{fixed} \spa \tilde{g}_s
\equiv \frac{g_s}{\epsilon}\ \mbox{fixed} \spa J_i\ \mbox{fixed}
\end{equation}
Here $E$ is the energy of the strings, $J_i$, $i=1,2,3$, are the
angular momenta for the five-sphere, $J = J_1+J_2$, $g_s$ is the
string coupling and $T_{\rm str}= R^2/(4\pi l_s^2) =
\sqrt{\lambda}/2$ is the string tension with $R$ being the AdS
radius and $l_s$ the string length. $\tilde{H}$ is the effective
Hamiltonian for the strings in the decoupling limit. We see that
both the string tension $T_{\rm str}$ and the string coupling $g_s$
go to zero in this limit.

As mentioned in the Introduction, we should now take a Penrose limit
of the $\ads_5\times S^5$ background and then consider the string
theory on the resulting pp-wave background. This gives the following
pp-wave background with 32 supersymmetries
\begin{equation}
\label{ppmet} \frac{ds^2}{\sqrt{\epsilon}} = - 4dx^+ dx^- - \mu^2
\sum_{I=3}^8 x^I x^I (dx^+)^2 + \sum_{i=1}^8 dx^i dx^i - 4 \mu x^2
dx^1 dx^+
\end{equation}
\begin{equation}
\label{ppF5} \frac{F_{(5)}}{\epsilon} = 2\mu dx^+ (dx^1 dx^2 dx^3
dx^4 + dx^5 dx^6 dx^7 dx^8 )
\end{equation}
This background was first found in \cite{Michelson:2002wa}%
\footnote{The pp-wave background \eqref{ppmet}-\eqref{ppF5} is
related to the maximally supersymmetric pp-wave background of
\cite{Blau:2001ne,Berenstein:2002jq} by a coordinate transformation
\cite{Michelson:2002wa,Bertolini:2002nr}. Even so, as we shall see
in the following, the physics of this pp-wave is rather different,
which basically origins in the fact that the coordinate
transformation between them depends on $x^+$, i.e. it is
time-dependent. See \cite{Bertolini:2002nr} for more comments on
this.}. It is important to note that in the pp-wave background
\eqref{ppmet}-\eqref{ppF5} the direction $x^1$ is an explicit
isometry of the pp-wave \cite{Michelson:2002wa,Bertolini:2002nr},
hence we call this background a pp-wave with a flat direction. The
resulting spectrum and level matching condition are given
by~\cite{Harmark:2006ta}
\begin{equation}
\label{fullstrspec}
\begin{array}{rcl} \ds \frac{l_s^2 p^+}{\sqrt{\epsilon}} H_{\rm lc} &=&\ds 2 f N_0 + \sum_{n\neq 0} \left[
(\omega_n+f) N_n + (\omega_n - f) M_n \right] + \sum_{n\in \Z}
\sum_{I=3}^8 \omega_n N_n^{(I)} \\[5mm] && \ds + \sum_{n\in \Z}
\left[ \sum_{b=1}^4 \left( \omega_n - \frac{1}{2} f \right)
F_n^{(b)} +\sum_{b=5}^8 \left( \omega_n + \frac{1}{2} f \right)
F_n^{(b)} \right]
\end{array}
\end{equation}
\begin{equation}
\label{levmat} \sum_{n\neq0} n \left[ N_n + M_n + \sum_{I=3}^8
N^{(I)}_n + \sum_{b=1}^8 F^{(b)}_n \right] = 0~,
\end{equation}
where we have defined $f = \mu l_s^2 p^+$ and $\omega_n = \sqrt{n^2
+ f^2}$. Here $N_n^{(I)}$, $I=3,...,8$ and $n\in Z$, are the number
operators for bosonic excitations for the six directions
$x^3,...,x^8$, while $N_n$, $n \in \Z$, and $M_n$, $n \neq 0$, are
the number operators for the two directions $x^1$ and $x^2$.
$F_n^{(b)}$, $b=1,...,8$ and $n\in \Z$, are the number operators for
the fermions. It is important to note that there is a vacuum for
each value of the momentum along the flat direction, and that
momentum is moreover dual to $J_1-J_2$. This is exactly as on the
gauge theory/spin chain side where and we have a vacuum for each
value of the total spin measured by $J_1-J_2$. Moreover we have a
pp-wave spectrum for which all states with $E=J$, $J=J_1+J_2$,
correspond to the string vacua, again as in the gauge theory side.

By then taking the large $\mu$ limit of the pp-wave
\begin{equation}
\label{pplim}  \epsilon \rightarrow 0 \spa \mu \rightarrow \infty
\spa \tilde{\mu} \equiv \mu \sqrt{\epsilon}\ \mbox{fixed} \spa
\tilde{H}_{\rm lc} \equiv \frac{H_{\rm lc}}{\epsilon}\ \mbox{fixed}
\spa \tilde{g}_s \equiv \frac{g_s}{\epsilon}\ \mbox{fixed} \spa
l_s,\ p^+ \ \mbox{fixed}
\end{equation}
which is an implementation of the limit \eqref{ouradslim}, we have
that the resulting spectrum, expressed in terms of gauge theory
quantities via the AdS/CFT correspondence, is given by
\begin{equation}
\label{spec2} \frac{1}{\tilde{\mu}} \tilde{H}_{\rm lc} =
\frac{2\pi^2 \tilde{\lambda}}{J^2} \sum_{n\neq 0} n^2 M_n \spa
\sum_{n\neq 0} n M_n = 0
\end{equation}
This precisely matches the gauge theory spectrum for large
$\tilde{\lambda}$ and $J$ in the decoupling limit \eqref{ourlim}.
Thus, we can match the spectrum of weakly coupled string theory with
weakly coupled gauge theory in the corresponding decoupling limits.

It is not difficult to show that from the matching of the spectra it
follows the matching of the Hagedorn
temperatures~\cite{Harmark:2006ta} which also on the string side is
given by equation \eqref{ourhag}~\footnote{The computation of the
string theory partition function and Hagedorn temperature can also
be done using the full spectrum \eqref{fullstrspec}. In this case we
get
\begin{equation}
\log Z ({a},{b},{\mu}) = \sum_{n=1}^{\infty} \frac{1}{n} \mbox{Tr}
\left( e^{-{a} n  H_{{\rm l.c.}}-{b}np^+}\right) \label{fab}
\end{equation}
where the parameters ${a}$ and ${b}$ can be viewed as inverse
temperature and chemical potential, respectively, for the pp-wave
strings. For related computations of the string theory partition
function and Hagedorn temperature in the presence of background
fields that play the role of chemical potentials for the
corresponding momenta see for example Ref.s \cite{Deo:1989bv,
Greene:2002cd}. From eq.n \eqref{fab} we get that the Hagedorn
temperature is defined by the following equation
\begin{equation}
{b} \sqrt{{a}} =  l_s^2 \zeth \sqrt{2\pi{\mu}} \label{hageab}
\end{equation}
In order to compare \eqref{hageab} with the gauge theory result
\eqref{ourhag} we have to express the parameters ${a}$ and ${b}$ in
terms of the gauge theory quantities and take the limit
\eqref{pplim}. It is easy to see that we get again the result
\eqref{ourhag}.}.

We have thus shown that the Hagedorn temperature of type IIB string
theory on $\ads_5 \times S^5$ in the decoupling limit
\eqref{ouradslim} matches with the Hagedorn/deconfinement
temperature \eqref{ourhag} computed in weakly coupled $\CN=4$ SYM in
the dual decoupling limit \eqref{ourlim}. This is done in the regime
of large $\tilde{\lambda}$. On the string side we obtained the
Hagedorn temperature by considering the large $\tilde{\lambda}$ and
$J$ limit corresponding to strings on the pp-wave background
\eqref{ppmet}-\eqref{ppF5} in the decoupling limit \eqref{pplim}.
The result means that in the sector of AdS/CFT defined by the
decoupling limits we can indeed show that the Hagedorn temperature
for type IIB string theory on the $\ads_5\times S^5$ background is
mapped to the Hagedorn/deconfinement temperature of weakly coupled
planar $\CN=4$ SYM on $\R \times S^3$. Thus we have direct evidence
that the confinement/deconfinement transition found in weakly
coupled planar $\CN=4$ SYM on $\R \times S^3$ is linked to a
Hagedorn transition of string theory on $\ads_5 \times S^5$, as
conjectured in
\cite{Witten:1998zw,Sundborg:1999ue,Polyakov:2001af,Aharony:2003sx}.

A similar computation for the spectrum and Hagedorn temperature can
be done for the situation dual to the decoupling limit \eqref{hlim}.
However there are interesting differences between the two cases
$\tilde h =0$ and $\tilde h \ne 0$. In the first case the vacuum of
the spin chain has an $L+1$ fold degeneracy since the states
$\tr(\textrm{sym}(Z^{L-M}X^M))$ all have the same energy for $0 \le
M \le L$.  In the case when an external magnetic field is present
this degeneracy is removed by the Zeeman term $\tilde h S_z$ and
$\tr(Z^L)$ becomes the unique vacuum. An analogous difference is
also present on the dual string theory side. It is possible to show
that the Hagedorn temperature for the string theory dual to the
gauge theory in the decoupling limit \eqref{hlim} is given by
\eqref{hhag}.

%%%%%%%%%%%%%%%%%%%%%%%%%%%%%%%%%%%%%%%%%%%%%%%%%%%%%%%%%%%%%%
\section*{Acknowledgments}

M.O. thank the organizers of the RTN workshop in Napoli where this
work was presented. TH would like to thank the Carlsberg Foundation
for support. The work of M.O. is supported in part by the European
Community's Human Potential Programme under contract
MRTN-CT-2004-005104 `Constituents, fundamental forces and symmetries
of the universe'.

%%%%%%%%%%%%%%%%%%%%%%%%%%%%%%%%%%%%%%%%%%%%%%%%%%%%%%%%%%

%The following two lines is for bibtex only:
%\bibliographystyle{utphys}
%\bibliography{mybib,bibrot,biblioniels}
%\bibliographystyle{../INPUT/utphys}
%\bibliography{../BIB/mybib,../BIB/bibrot,../BIB/biblioniels}

\begin{thebibliography}{10}

\bibitem{Harmark:2006di}
T.~Harmark and M.~Orselli, ``Quantum mechanical sectors in thermal
{$\CN = 4$}
  super {Yang-Mills} on {$\R \times S^3$},'' {\em Nucl. Phys.} {\bf B757}
  (2006) 117--145,
\href{http://www.arXiv.org/abs/hep-th/0605234}{{\tt
hep-th/0605234}}.
%%CITATION = HEP-TH 0605234;%%.

\bibitem{Maldacena:1997re}
J.~M. Maldacena, ``The large {$N$} limit of superconformal field
theories and
  supergravity,'' {\em Adv. Theor. Math. Phys.} {\bf 2} (1998) 231--252,
\href{http://www.arXiv.org/abs/hep-th/9711200}{{\tt
hep-th/9711200}}.
%%CITATION = HEP-TH 9711200;%%.

\bibitem{Gubser:1998bc}
S.~S. Gubser, I.~R. Klebanov, and A.~M. Polyakov, ``Gauge theory
correlators
  from noncritical string theory,'' {\em Phys. Lett.} {\bf B428} (1998) 105,
\href{http://www.arXiv.org/abs/hep-th/9802109}{{\tt
hep-th/9802109}}.
%%CITATION = PHLTA,B428,105;%%.

\bibitem{Witten:1998qj}
E.~Witten, ``{Anti-de Sitter} space and holography,'' {\em Adv.
Theor. Math.
  Phys.} {\bf 2} (1998) 253,
\href{http://www.arXiv.org/abs/hep-th/9802150}{{\tt
hep-th/9802150}}.
%%CITATION = 00203,2,253;%%.

\bibitem{Erickson:2000af}
J.~K. Erickson, G.~W. Semenoff, and K.~Zarembo, ``Wilson loops in
{$\CN = 4$}
  supersymmetric {Yang-Mills} theory,'' {\em Nucl. Phys.} {\bf B582} (2000)
  155--175,
\href{http://www.arXiv.org/abs/hep-th/0003055}{{\tt
hep-th/0003055}}.
%%CITATION = HEP-TH 0003055;%%.

\bibitem{Drukker:2000rr}
N.~Drukker and D.~J. Gross, ``An exact prediction of {$\CN = 4$}
{SUSYM} theory
  for string theory,'' {\em J. Math. Phys.} {\bf 42} (2001) 2896--2914,
\href{http://www.arXiv.org/abs/hep-th/0010274}{{\tt
hep-th/0010274}}.
%%CITATION = HEP-TH 0010274;%%.

\bibitem{Berenstein:2002jq}
D.~Berenstein, J.~M. Maldacena, and H.~Nastase, ``Strings in flat
space and pp
  waves from {$\CN = 4$} super {Yang Mills},'' {\em JHEP} {\bf 04} (2002) 013,
\href{http://www.arXiv.org/abs/hep-th/0202021}{{\tt
hep-th/0202021}}.
%%CITATION = HEP-TH 0202021;%%.

\bibitem{Harmark:2006ta}
T.~Harmark and M.~Orselli, ``Matching the {Hagedorn} temperature in
  {AdS/CFT},'' {\em Phys. Rev.} {\bf D74} (2006) 126009,
\href{http://www.arXiv.org/abs/hep-th/0608115}{{\tt
hep-th/0608115}}.
%%CITATION = HEP-TH 0608115;%%.

\bibitem{Harmark:2006ie}
T.~Harmark, K.~R. Kristjansson, and M.~Orselli, ``Magnetic
{Heisenberg-chain} /
  pp-wave correspondence,'' {\em JHEP} {\bf 02} (2007) 085,
\href{http://www.arXiv.org/abs/hep-th/0611242}{{\tt
hep-th/0611242}}.
%%CITATION = HEP-TH/0611242;%%.

\bibitem{Witten:1998zw}
E.~Witten, ``{Anti-de Sitter space}, thermal phase transition, and
confinement
  in gauge theories,'' {\em Adv. Theor. Math. Phys.} {\bf 2} (1998) 505,
\href{http://www.arXiv.org/abs/hep-th/9803131}{{\tt
hep-th/9803131}}.
%%CITATION = 00203,2,505;%%.

\bibitem{Sundborg:1999ue}
B.~Sundborg, ``The {Hagedorn} transition, deconfinement and {$\CN =
4$ SYM}
  theory,'' {\em Nucl. Phys.} {\bf B573} (2000) 349--363,
\href{http://www.arXiv.org/abs/hep-th/9908001}{{\tt
hep-th/9908001}}.
%%CITATION = HEP-TH 9908001;%%.

\bibitem{Polyakov:2001af}
A.~M. Polyakov, ``Gauge fields and space-time,'' {\em Int. J. Mod.
Phys.} {\bf
  A17S1} (2002) 119--136,
\href{http://www.arXiv.org/abs/hep-th/0110196}{{\tt
hep-th/0110196}}.
%%CITATION = HEP-TH 0110196;%%.

\bibitem{Aharony:2003sx}
O.~Aharony, J.~Marsano, S.~Minwalla, K.~Papadodimas, and
M.~Van~Raamsdonk,
  ``The {H}agedorn/deconfinement phase transition in weakly coupled large {$N$}
  gauge theories,'' {\em Adv. Theor. Math. Phys.} {\bf 8} (2004) 603--696,
\href{http://www.arXiv.org/abs/hep-th/0310285}{{\tt
hep-th/0310285}}.
%%CITATION = HEP-TH 0310285;%%.

\bibitem{Bertolini:2002nr}
M.~Bertolini, J.~de~Boer, T.~Harmark, E.~Imeroni, and N.~A. Obers,
``Gauge
  theory description of compactified pp-waves,'' {\em JHEP} {\bf 01} (2003)
  016,
\href{http://www.arXiv.org/abs/hep-th/0209201}{{\tt
hep-th/0209201}}.
%%CITATION = HEP-TH 0209201;%%.

\bibitem{Greene:2002cd}
B.~R. Greene, K.~Schalm, and G.~Shiu, ``On the {Hagedorn} behaviour
of
  {pp-wave} strings and {$\CN = 4$} {SYM} theory at finite {R-charge}
  density,'' {\em Nucl. Phys.} {\bf B652} (2003) 105--126,
\href{http://www.arXiv.org/abs/hep-th/0208163}{{\tt
hep-th/0208163}}.
%%CITATION = HEP-TH 0208163;%%.




\bibitem{PandoZayas:2002hh}
L.~A. Pando~Zayas and D.~Vaman, ``Strings in {RR} plane wave
background at
  finite temperature,'' {\em Phys. Rev.} {\bf D67} (2003) 106006,
\href{http://www.arXiv.org/abs/hep-th/0208066}{{\tt
hep-th/0208066}}.
%%CITATION = HEP-TH 0208066;%%.
%\bibitem{Sugawara:2002rs}
Y.~Sugawara, ``Thermal amplitudes in {DLCQ} superstrings on
{pp-waves},'' {\em
  Nucl. Phys.} {\bf B650} (2003) 75--113,
\href{http://www.arXiv.org/abs/hep-th/0209145}{{\tt
hep-th/0209145}}.
%%CITATION = HEP-TH 0209145;%%.
%\bibitem{Brower:2002zx}
R.~C. Brower, D.~A. Lowe, and C.-I. Tan, ``Hagedorn transition for
strings on
  {pp-waves} and tori with chemical potentials,'' {\em Nucl. Phys.} {\bf B652}
  (2003) 127--141,
\href{http://www.arXiv.org/abs/hep-th/0211201}{{\tt
hep-th/0211201}}.
%%CITATION = HEP-TH 0211201;%%.
%\bibitem{Sugawara:2003qc}
Y.~Sugawara, ``Thermal partition function of superstring on
compactified
  {pp-wave},'' {\em Nucl. Phys.} {\bf B661} (2003) 191--208,
\href{http://www.arXiv.org/abs/hep-th/0301035}{{\tt
hep-th/0301035}}.
%%CITATION = HEP-TH 0301035;%%.
%\bibitem{Grignani:2003cs}
G.~Grignani, M.~Orselli, G.~W. Semenoff, and D.~Trancanelli, ``The
superstring
  {Hagedorn} temperature in a {pp-wave} background,'' {\em JHEP} {\bf 06}
  (2003) 006,
\href{http://www.arXiv.org/abs/hep-th/0301186}{{\tt
hep-th/0301186}}.
%%CITATION = HEP-TH 0301186;%%.
%\bibitem{Hyun:2003ks}
S.-j. Hyun, J.-D. Park, and S.-H. Yi, ``Thermodynamic behavior of
{IIA} string
  theory on a {pp-wave},'' {\em JHEP} {\bf 11} (2003) 006,
\href{http://www.arXiv.org/abs/hep-th/0304239}{{\tt
hep-th/0304239}}.
%%CITATION = HEP-TH 0304239;%%.
%\bibitem{Bigazzi:2003jk}
F.~Bigazzi and A.~L. Cotrone, ``On zero-point energy, stability and
{Hagedorn}
  behavior of type {IIB} strings on {pp-waves},'' {\em JHEP} {\bf 08} (2003)
  052,
\href{http://www.arXiv.org/abs/hep-th/0306102}{{\tt
hep-th/0306102}}.
%%CITATION = HEP-TH 0306102;%%.










\bibitem{AlvarezGaume:2006jg}
L.~Alvarez-Gaume, P.~Basu, M.~Marino, and S.~R. Wadia, ``Blackhole /
string
  transition for the small {Schwarzschild} blackhole of {$\mbox{AdS}_5 \times
  S^5$} and critical unitary matrix models,'' {\em Eur. Phys. J.} {\bf C48}
  (2006) 647--665,
\href{http://www.arXiv.org/abs/hep-th/0605041}{{\tt
hep-th/0605041}}.
%%CITATION = HEP-TH/0605041;%%.
%
%\bibitem{Hollowood:2006xb}
T.~Hollowood, S.~P. Kumar, and A.~Naqvi, ``Instabilities of the
small black
  hole: A view from {$\CN = 4$} {SYM},'' {\em JHEP} {\bf 01} (2007) 001,
\href{http://www.arXiv.org/abs/hep-th/0607111}{{\tt
hep-th/0607111}}.
%%CITATION = HEP-TH/0607111;%%.
%
%\bibitem{Hartnoll:2006pj}
S.~A. Hartnoll and S.~P. Kumar, ``Thermal {$\CN = 4$} {SYM} theory
as a {2D}
  {Coulomb} gas,'' {\em Phys. Rev.} {\bf D76} (2007) 026005,
\href{http://www.arXiv.org/abs/hep-th/0610103}{{\tt
hep-th/0610103}}.
%%CITATION = HEP-TH/0610103;%%.
%
%\bibitem{Festuccia:2006sa}
G.~Festuccia and H.~Liu, ``The arrow of time, black holes, and
quantum mixing
  of large {N} {Yang-Mills} theories,''
\href{http://www.arXiv.org/abs/hep-th/0611098}{{\tt
hep-th/0611098}}.
%%CITATION = HEP-TH 0611098;%%.

\bibitem{Harmark:1999xt}
T.~Harmark and N.~A. Obers, ``Thermodynamics of spinning branes and
their dual
  field theories,'' {\em JHEP} {\bf 01} (2000) 008,
\href{http://www.arXiv.org/abs/hep-th/9910036}{{\tt
hep-th/9910036}}.
%%CITATION = JHEPA,0001,008;%%.

\bibitem{Minahan:2002ve}
J.~A. Minahan and K.~Zarembo, ``The {Bethe-ansatz} for {$\CN = 4$}
super
  {Yang-Mills},'' {\em JHEP} {\bf 03} (2003) 013,
\href{http://www.arXiv.org/abs/hep-th/0212208}{{\tt
hep-th/0212208}}.
%%CITATION = HEP-TH 0212208;%%.

\bibitem{Michelson:2002wa}
J.~Michelson, ``(twisted) toroidal compactification of pp-waves,''
{\em Phys.
  Rev.} {\bf D66} (2002) 066002,
\href{http://www.arXiv.org/abs/hep-th/0203140}{{\tt
hep-th/0203140}}.
%%CITATION = HEP-TH 0203140;%%.

\bibitem{Blau:2001ne}
M.~Blau, J.~Figueroa-O'Farrill, C.~Hull, and G.~Papadopoulos, ``A
new maximally
  supersymmetric background of {IIB} superstring theory,'' {\em JHEP} {\bf 01}
  (2002) 047,
\href{http://www.arXiv.org/abs/hep-th/0110242}{{\tt
hep-th/0110242}}.
%%CITATION = HEP-TH 0110242;%%.



\bibitem{Deo:1989bv}
N.~Deo, S.~Jain, and C.-I. Tan, ``String statistical mechanics above
Hagedorn
  energy density,'' {\em Phys. Rev.} {\bf D40} (1989)
2626.
%%CITATION = PHRVA,D40,2626;%%.
%
%\bibitem{Brower:1998an}
R.~C. Brower, J.~McGreevy, and C.~I. Tan, ``Stringy model for QCD at
finite
  density and generalized Hagedorn temperature,''
\href{http://www.arXiv.org/abs/hep-ph/9907258}{{\tt
hep-ph/9907258}}.
%%CITATION = HEP-PH/9907258;%%.
%
%\bibitem{Grignani:2001hb}
G.~Grignani, M.~Orselli, and G.~W. Semenoff, ``Matrix strings in a
B-field,''
  {\em JHEP} {\bf 07} (2001) 004,
\href{http://www.arXiv.org/abs/hep-th/0104112}{{\tt
hep-th/0104112}}.
%%CITATION = HEP-TH/0104112;%%.
%
%\bibitem{Grignani:2001ik}
G.~Grignani, M.~Orselli, and G.~W. Semenoff, ``The target space
dependence of
  the Hagedorn temperature,'' {\em JHEP} {\bf 11} (2001) 058,
\href{http://www.arXiv.org/abs/hep-th/0110152}{{\tt
hep-th/0110152}}.
%%CITATION = HEP-TH/0110152;%%.
%
%\bibitem{DeRisi:2002gt}
G.~De~Risi, G.~Grignani, and M.~Orselli, ``Space / time
noncommutativity in
  string theories without background electric field,'' {\em JHEP} {\bf 12}
  (2002) 031,
\href{http://www.arXiv.org/abs/hep-th/0211056}{{\tt
hep-th/0211056}}.
%%CITATION = HEP-TH/0211056;%%.






\end{thebibliography}

\providecommand{\href}[2]{#2}\begingroup\raggedright\endgroup

\end{document}